\documentclass[11pt, conference]{IEEEtran}
\IEEEoverridecommandlockouts
\usepackage{hyperref}
\usepackage[a4paper,left=0.625in,right=0.625in,top=0.75in,bottom=1.69in]{geometry}

\usepackage{amsmath} 
\usepackage{amssymb}  
\usepackage{mathtools}
\usepackage{upgreek}
\usepackage{color,soul}
\usepackage{lipsum}
\usepackage{float}
\usepackage{balance}
\usepackage{url}
\usepackage{xr}
\usepackage{algpseudocode}
\usepackage{xcolor}
\usepackage{breqn}

\usepackage{booktabs}
\usepackage{cite}
\usepackage{graphicx}
\usepackage{graphics} 
\usepackage{subcaption} 
\usepackage{epsfig} 
\usepackage{gensymb}

\usepackage[linesnumbered,ruled,vlined]{algorithm2e}
\algnewcommand{\algorithmicand}{\textbf{ and }}
\algnewcommand{\algorithmicor}{\textbf{ or }}
\algnewcommand{\algorithmicnot}{\textbf{ not }}
\algnewcommand{\OR}{\algorithmicor}
\algnewcommand{\AND}{\algorithmicand}
\algnewcommand{\NOT}{\algorithmicnot}

\usepackage[most]{tcolorbox}
\usepackage{amssymb}
\usepackage{amsthm}
\usepackage{placeins}

\algdef{SE}[VARIABLES]{Variables}{EndVariables}
   {\algorithmicvariables}
   {\algorithmicend\ \algorithmicvariables}
\algnewcommand{\algorithmicvariables}{\textbf{global variables}}
\algdef{SE}[VARIABLESDEF]{VariablesDef}{EndVariablesDef}
   {\algorithmicvariablesdef}
   {\algorithmicend\ \algorithmicvariablesdef}
\algnewcommand{\algorithmicvariablesdef}{\textbf{variables definition}}

\title{IntAgent: NWDAF-Based Intent LLM Agent Towards Advanced Next Generation Networks}

	\author{
	\IEEEauthorblockN{ Abdelrahman Soliman\IEEEauthorrefmark{1}, Ahmed Refaey\IEEEauthorrefmark{1}, Aiman Erbad \IEEEauthorrefmark{2}, and Amr Mohamed\IEEEauthorrefmark{2}}

	\IEEEauthorblockA{\IEEEauthorrefmark{1} University of Guelph, Guelph, Ontario, Canada.}
    \IEEEauthorblockA{\IEEEauthorrefmark{2} Qatar University, Doha, Qatar.}}

\begin{document}

\maketitle

\begin{abstract}
Intent-based networks (IBNs) are gaining prominence as an innovative technology that automates network operations through high-level request statements, defining what the network should achieve. In this work, we introduce IntAgent, an intelligent intent LLM agent that integrates NWDAF analytics and tools to fulfill the network operator's intents. Unlike previous approaches, we develop an intent tools engine directly within the NWDAF analytics engine, allowing our agent to utilize live network analytics to inform its reasoning and tool selection. We offer an enriched, 3GPP-compliant data source that enhances the dynamic, context-aware fulfillment of network operator goals, along with an MCP tools server for scheduling, monitoring, and analytics tools. We demonstrate the efficacy of our framework through two practical use cases: ML-based traffic prediction and scheduled policy enforcement, which validate IntAgent's ability to autonomously fulfill complex network intents.
\end{abstract}

\begin{IEEEkeywords}
	\makeatletter
	\def\RemoveSpaces#1{\zap@space#1 \@empty}
	\makeatother 
Core Network,
Next Generation Network,
Large Language Models,
Intent Management,
Intelligent Networks,
Closed Loop,
\end{IEEEkeywords}

\section{INTRODUCTION}
\label{sec:intro}

Rapid surge in network traffic and exponential growth of connected devices are driving the development of next-generation networks (NGNs), including 6G and beyond \cite{saravanan2025architecture}. 5G has already transformed network requirements and technologies, particularly in management and design. In this context, artificial intelligence (AI) and machine learning (ML) have emerged as key enablers for intent-based networking, automation, and optimization. To support this vision, the 3rd Generation Partnership Project (3GPP) introduced the Network Data Analytics Function (NWDAF), the first standardized integration of AI into the core network \cite{he2023nwdaf}. NWDAF acts as a centralized analytics engine, collecting data from network functions, user equipment, and management systems \cite{9143579} to enable intelligent automation and improve operations. Traditional network management is increasingly inadequate, relieving on manual processes that are slow, error-prone, and resource-intensive \cite{gonzalez2025extending}. Intent-based networking (IBN) offers a solution by automating operations through high-level statements of network intent, shifting the focus from configuration to goals. This approach improves agility and efficiency, allowing the network to adapt dynamically with minimal human input. However, IBN faces challenges, including ensuring security and compliance in automated systems \cite{wei2020intent} and accurately translating high-level intentions into specific configurations, which requires advanced algorithms.

Large Language Models (LLMs) are gaining prominence as a groundbreaking technology for automating complex systems. Their core roles include understanding natural language, generating code, and reasoning logically. This is evolving into an agentic function, in which LLMs can not only process information, but also plan and execute actions on their own to achieve a desired goal \cite{bandi2025rise}. In IBN, this role is core; an LLM can act as a cognitive engine, translating a network operator's high-level intent into specific device configurations. In order to achieve trustworthy automation, LLMs must be fine-tuned with domain-specific networking knowledge, provided real-time network telemetry for context, and granted secure, controlled access to network management interfaces.


Recent research increasingly explores the use of LLMs for intent-based networking (IBN) management, from intent translation to autonomous agentic systems. One focus is intent translation, where LLMs convert natural language requests into structured network configurations. For instance, \cite{manias2024towards,gonzalez2025extending} extract and classify high-level user intents into precise actions for 5G and B5G networks, while \cite{asisi2025intent} applies a similar approach for RAN management. However, these studies often lack deep integration with network analytics functions such as NWDAF for data-driven real-time decisions. Another line of research proposes agentic frameworks that plan and execute complex tasks. Works by \cite{mekrache2025oss,wang2025intent,brodimas2025intent} develop single and multi-agent systems to orchestrate API calls or use Infrastructure-as-Code tools. Similarly, \cite{guo2025intent,liu2025lameta} combines LLM-based intent handling with ML paradigms like deep reinforcement learning for closed-loop resource allocation. Despite their automation, these systems assume preexisting tools and abstract network states. They rarely address tool creation based on standardized analytics or dynamic use of live network data, as highlighted by \cite{ardestani2025towards} in the NWDAF pipeline.

In contrast, our work introduces IntAgent, an intelligent LLM agent that uniquely bridges the gap between high-level intent management and standardized real-time network intelligence. Unlike previous approaches, we develop an intent tools engine directly within the NWDAF analytics engine, allowing our agent to utilize live network analytics to inform its reasoning and tool selection. By extending NWDAF's event exposure capabilities, we provide IntAgent with a richer, 3GPP-compliant data source, enabling a more dynamic and context-aware approach to fulfilling network operator intents.

The contributions of this paper can be summarized as follows:
\begin{itemize}
	  \item We extend NWDAF notification and event exposure for other network functions, allowing for more extended data collection and analytics following the 3GPP standards.
        \item We developed the intent tools engine in the analytics engine, which will provide a Model Context Protocol (MCP) server that serves the scheduling, monitoring, and analytics tools.
        \item We propose IntAgent, an intelligent intent LLM agent that incorporates NWDAF analytics and tools to fulfill the network operator's intents. 
        \item We provide findings and insights on two tested use cases, which verify the functionality of the agent. 
  
\end{itemize}  

The rest of the paper is organized as follows: Section II presents an overview of the architecture of our system. Section III provides details about the proposed agent and its algorithm. Section IV presents and discusses the experiments and findings. Finally, Section V provides conclusions and future work.

\section{SYSTEM ARCHITECTURE}
\begin{figure*}[h] 
    \centering
    \includegraphics[width=1\linewidth,height=0.3\textheight,keepaspectratio]{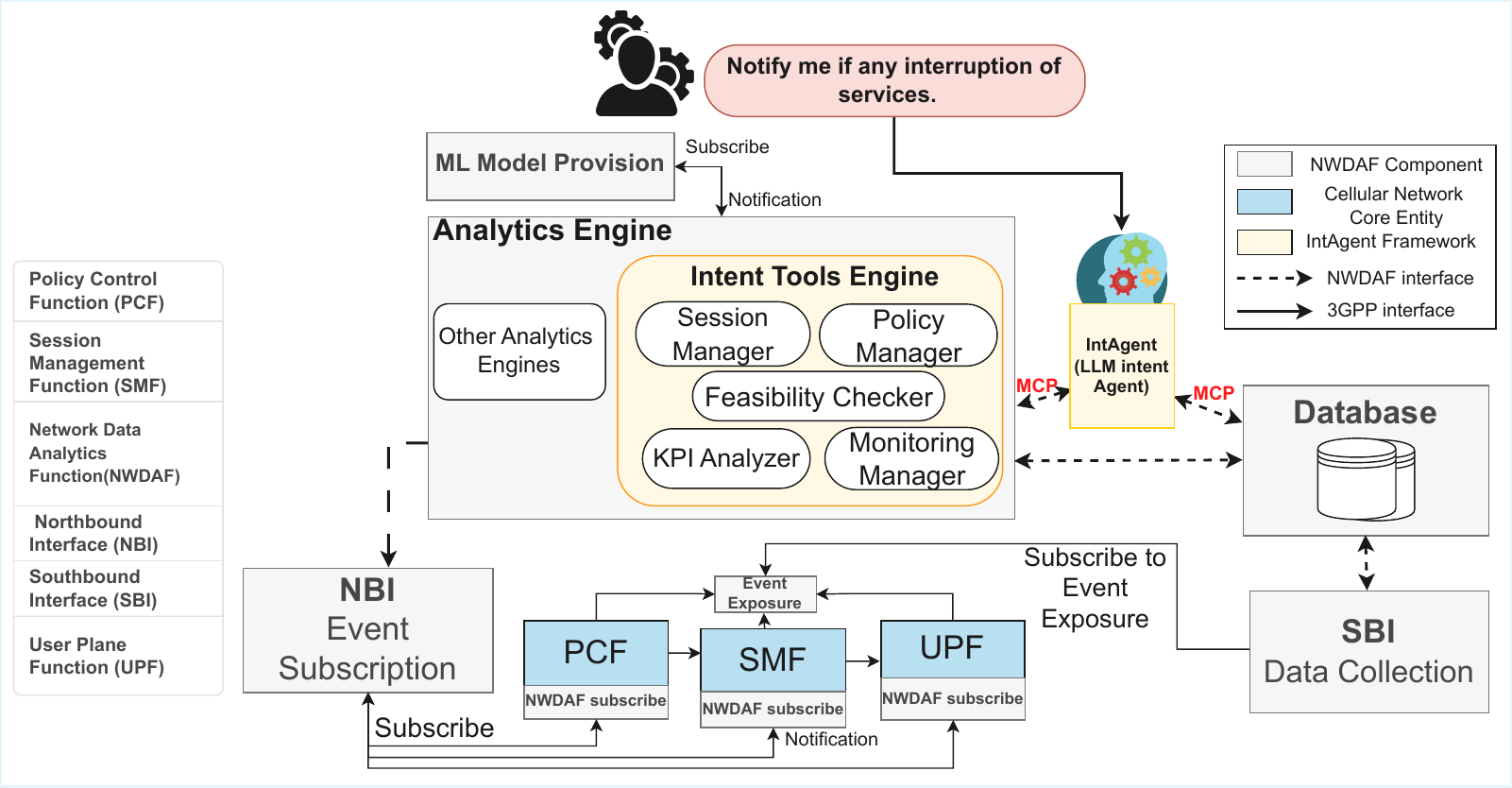}
    \caption{System model showing IntAgent integrated with NWDAF to enable intent-driven management through tools, event subscription, and data collection.}

    \label{fig:arch}
\end{figure*}

Figure \ref{fig:arch} presents the proposed system architecture. The primary goal is to transform the network operator's intentions and issues regarding the network into tangible actions and results. The LLM agent will use the analytics provided to understand the context and make use of intent tools to generate actionable insights and policies. 

\subsection{Overview of the Architecture}
The system comprises three principal components: the cellular core network entities, the NWDAF network function, and ultimately, the IntAgent Framework. We use core NFs such as PCF (Policy Control Function), SMF (Session Management Function), and UPF (User Plane which is responsible for sending periodic data for the NWDAF function through SBI (Service Based Interface) \cite{ardestani2025towards}. It also subscribes to the Northbound Interface (NBI) to receive recommendations and notifications from both existing analytics functions in the Analytics engine and configurations from IntAgent tools. 

\subsection{NWDAF Components}
NWDAF is typically made up of four key parts: data collection, analytics engine, model management, and exposure interfaces that provide insight to other network functions.
In this work, we build on the open source efforts detailed in \cite{mekrache2023combining,ardestani2025towards}, where a closed-loop NWDAF implementation compliant with 3GPP standards was presented, utilizing established 5G core network solutions such as Open5GS \cite{open5gs2025}. Raw data is gathered through the Event Exposure Service (EES) by collecting data via SBI and storing it in the MongoDB database. In \cite{ardestani2025towards}, the authors used EES for the UPF function specifically to collect traffic data. In contrast, our study expands the use of EES to include data and metrics for not just the UPF, but also the PCF and SMF. In addition, they implemented an ML model management system for model lifecycle tasks, such as saving and progress tracking.

\subsection{Intent Tools Engine}
The analytics engine is an intrinsic component of NWDAF, where it provides the available analytics for NFs. For example, authors in \cite{ardestani2025towards} created a bot detection engine to flag anomalous user equipment (UE) through ML analysis from collected UPF traffic data and release its PDU session automatically. Rather than crafting specific scenarios, we designed a versatile set of intention tools that leverage data from the NWDAF database. The LLM agent will call these tools to retrieve insights and actions that depend on the given intent context. 

The Feasibility Checker serves as the initial checkpoint, utilizing constraint evaluation to assess which actions are possible. The KPI Analyzer conducts statistical assessments on integrated telemetry data to identify key performance indicator (KPI) pattern shifts preceding service quality decline. The Monitoring Manager operates at the same time to manage time-based operations through scheduling and monitoring of given metrics and actions. The policy manager enforces policies through a process that compares new policy implementations with existing governance standards. The Session Manager enables detailed management of session lifecycles through its validation system, which checks QoS changes and state transitions against SMF and UPF standards.     

\section{PROPOSED AGENT}
In this section, we discuss the detailed structure of our proposed IntAgent and the workflow with the tools. 

\subsection{High-Level Architecture}

\begin{figure}[h]
	\centering
	\includegraphics[width=0.62\linewidth]{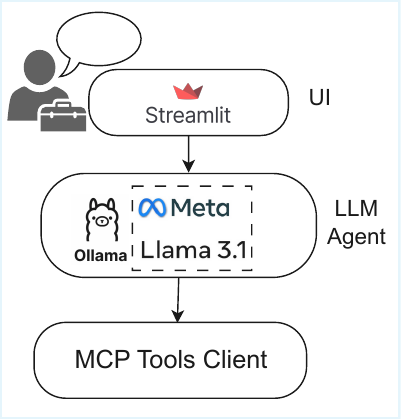}
	\caption{IntAgent architecture.}
	\label{fig:intagent}
\end{figure}
As depicted in Figure \ref{fig:intagent}, the system operator will access the user interface through a website hosted by Streamlit. This site includes an input box for user intent entry. Upon submission, it will be sent to Ollama, which is an application to host and run local LLMs on the machine. In our paper, we employed Meta Llama 3.1, a robust model with 8 billion parameters that features long context window support and an MCP tools calling capability. 

\subsection{Core Components of IntAgent}
The system's core components are the LLM and the toolset. LLM acts as both the brain and orchestrator, progressing through multiple steps to achieve the intended goal and requiring tools to execute actions and configurations.
\subsubsection{LLM prompt}
In order to direct the LLM model towards concentrating on networking knowledge and adhering to a sequence of steps to achieve the intended goal, we crafted a tailored system prompt. This prompt is complemented by a few-shot examples, offering guidance to the LLM model. Below provides a detailed overview of the prompt. The LLM model's response is restricted to three specific formats: thought, tool call, and final answer. The 'thought' format is employed for reasoning and observations, 'tool call' is used to invoke a particular tool, and the 'final answer' format compiles a summary and presents the results as a conclusive response.

\begin{tcolorbox}[colback=gray!20!white, colframe=gray, title=\textbf{InAgent LLM System Prompt}]

You are IntAgent, an advanced intent
agent for 5G network operations. Use tools to answer user intent.
Always answer in JSON using one of the
keys thought, tool_call, or
final_answer.
1. Call List Tools to
Understand which tool to use
2. Plan First
3. Discover Context
4. Feasibility Check
5. Follow Your Plan
6. Explain Observations
7. Finalize Clearly
\end{tcolorbox}

\subsubsection{Tools Set}
We enabled the agent to access NWDAF data and interact with real-world actions through MCP \cite{hou2025model}, which functions as an industry-standard protocol for connecting tools with AI agents powered by LLMs. The MCP system employs a universal client-server architecture, which unifies the connection process instead of needing separate integration work for each connection, as traditional APIs do. 

\begin{figure}[h]
	\centering
	\includegraphics[width=1\linewidth]{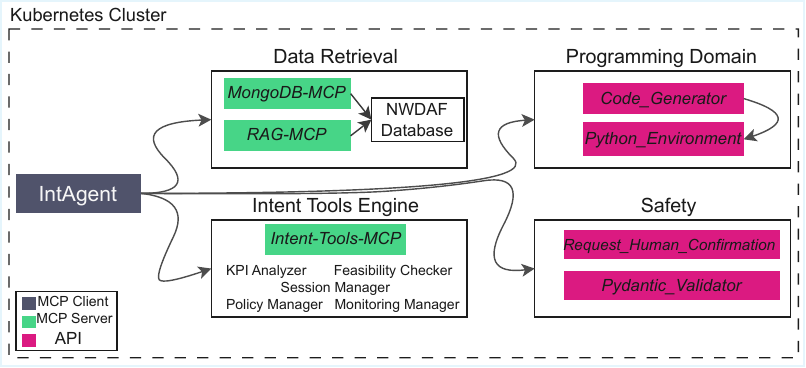}
	\caption{IntAgent tools connection schema.}
	\label{fig:tools}
\end{figure}

We connected our LLM as an MCP client, with multiple tools as shown in Figure \ref{fig:tools}. We categorized the tools into four groups: Data retrieval, which is responsible for extracting needed context from the NWDAF database; intent tools, which we discussed in the previous section; programming tools, which are responsible for generating and executing Python code using DeepSeek-R1 LLM model; and finally, safety tools, to verify the parameters calling and request human confirmation before taking any change action that could impact the network state.

\subsection{Agentic Loop}
The Agentic Loop stands as the core operational system of IntAgent, which enables LLMs to function as self-directed agents that perform multiple reasoning steps and complete tasks \cite{bandi2025rise}. As illustrated in Algorithm \ref{alg:agentic-loop}, the loop starts by creating a plan through integrating information from both the system prompt and user intent. The agent performs all steps through tool calls, which generate observations that get processed to update its current context. A critical feature of the loop is critical thinking, which ensures robust and safe behavior through four mechanisms which include \textbf{assumptions blocking} that stops dangerous actions based on false assumptions and \textbf{goal tracking} which monitors progress toward objectives and main intent and \textbf{structured validation} which checks tool calls and parameters before execution to prevent invalid operations and \textbf{safety gate-keeping} which asks for human approval when critical actions occur. 

\begin{algorithm}
\caption{Agentic Loop pseudocode}
\label{alg:agentic-loop}
\DontPrintSemicolon
\SetKwFunction{RunAgent}{RunAgent}
\RunAgent{intent}{;
  DiscoverMCPtools()\;
  context $\gets$ BuildSystemPrompt(intent)\;

  \While{not StopRequested() \textbf{and} not GoalAchieved(context, intent)}{
    
    \tcc{--- Plan ---}
    plan $\gets$ GenerateThought(context, intent)\;
    RecordHistory("thought", plan)\;

    \ForEach{step in plan.steps}{
      
      \tcc{--- Act ---}
      tool\_call $\gets$ PrepareToolCall(step, context)\;
      obs $\gets$ ExecuteTool(tool\_call)\;
      
      \tcc{--- Observe ---}
      \If{SafetyViolation(obs)}{
        RequestHumanConfirmation(); HaltIfUnapproved()\;
      }
      
      \tcc{--- Critical-Thinking ---}
      exp $\gets$ InterpretObservation(obs, intent)\;
      UpdateContext(context, obs, exp)\;
    }

    final $\gets$ SummarizeContext(context)\;
    \If{GoalAchieved(final, intent)}{\Return final}
  }
}
\end{algorithm}

\section{EXPERIMENTS AND ANALYSIS}
To validate the IntAgent's capabilities, we will introduce two distinct test scenarios: one focused on data comprehension and producing machine learning analytics, and the other concentrated on scheduling and monitoring tasks. 

\subsection{Deployment setup}\vspace{-0.5em}

An implementation of the core network was carried out to serve as a demonstration model for the network infrastructure using the deployment of an Open5GS Kubernetes cluster. We also implemented the updated NWDAF with intent tools and IntAgent. In our tests involving UE and RAN, we used the UERANSIM simulator to establish connections for 10 UEs that generated random traffic. The hardware configuration used was an i7-13700K CPU, 32 GB of RAM, and an NVIDIA RTX 3050 graphics card. The LLM model temperature was set to 0.1 to achieve near-deterministic results by reducing randomness.
To further validate the system, we monitored latency, throughput, and signaling overhead across multiple sessions, confirming stable operation under simulated traffic bursts. This experimental setup demonstrates the feasibility of integrating intent-based orchestration with AI-driven analytics directly into the core network, paving the way for scalable, autonomous management solutions.

\subsection{Use case 1}
\begin{tcolorbox}[colback=blue!20!white, colframe=blue, title=Use Case 1: Machine learning intent]
Want to predict memory utilization \% for the internet slice based on 500 recent values.
\end{tcolorbox}

\begin{figure}[h]
	\centering
	\includegraphics[width=1\linewidth]{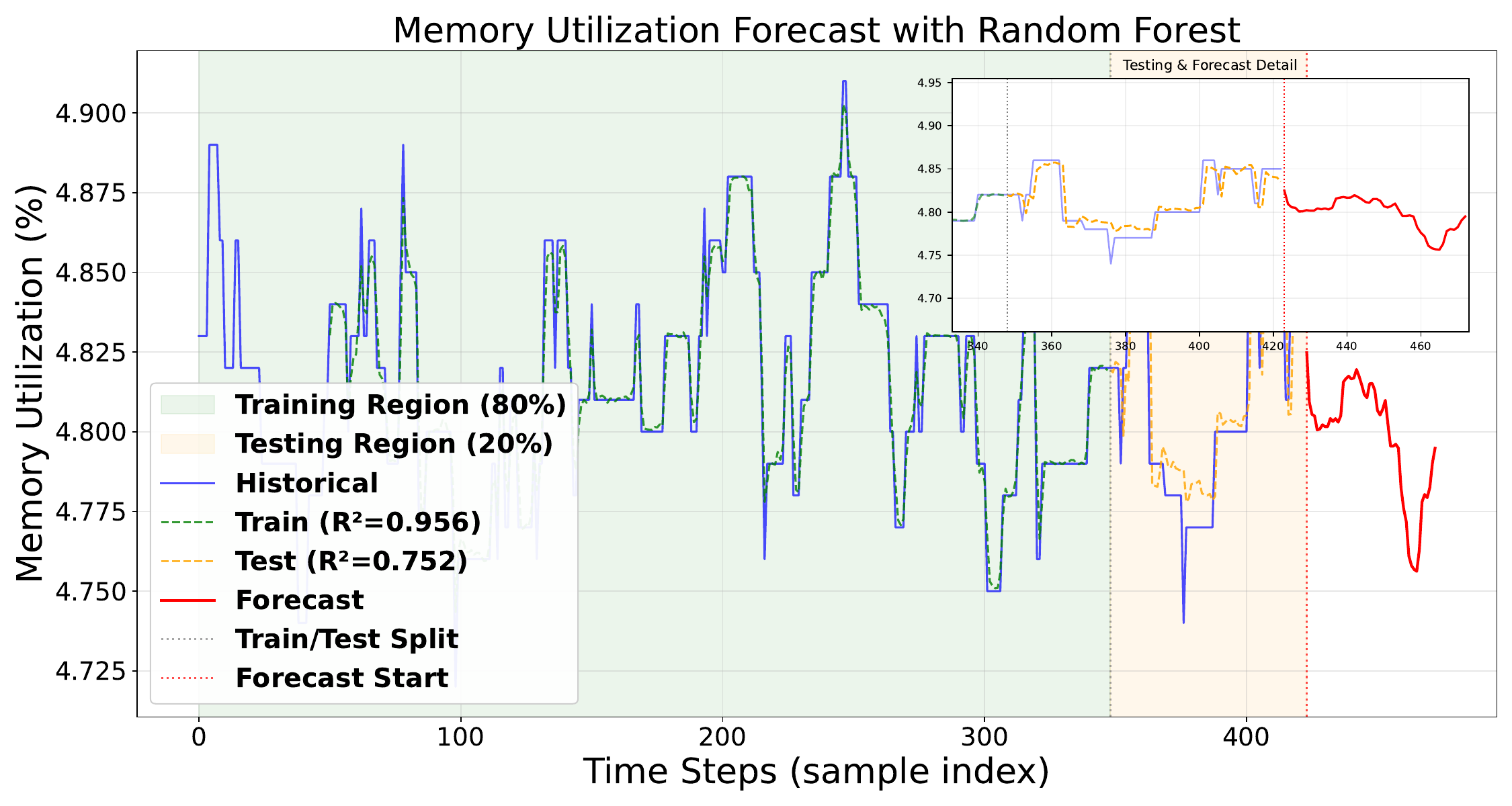}
	\caption{Plot obtained from IntAgent final answer for Intent 1.}
	\label{fig:res1}
\end{figure}


The first use case focuses on requesting a future prediction using machine learning. The agent started by planning, extracting the required data, generating, and executing Python code. The agent ultimately produced a diagram, shown in Figure \ref{fig:res1}, which illustrates the successful training of a Random Forest regressor model. The agent’s final answer, highlighted in the red box below, confirms that the intended task was achieved successfully.
In addition to model training, the agent validated its output by comparing predicted values against ground truth samples, ensuring accuracy and reliability. This use case demonstrates not only the agent’s capability for automated reasoning and execution but also its potential for scalable deployment in core network decision-support scenarios.

\begin{tcolorbox}[colback=red!10!white, colframe=red, title= \textbf{Use Case 1: Final Answer}]
\small
Your goal of predicting memory utilization for the internet slice based on 500 recent values has been successfully achieved using a Random Forest
Regressor model with a sliding window approach. The methodology involved loading and filtering data, removing outliers, preparing features,
training and testing the model, and finally making predictions on existing and future data. Notably, the R-squared value for the test set is 0.7520,
indicating a moderate level of accuracy in predicting memory utilization.  The forecast plot has been saved to
'/tmp/enhanced_forecast_1758468715.png' for your reference.
\end{tcolorbox}

\subsection{Use case 2}
\begin{figure}[h]
	\centering
	\includegraphics[width=1\linewidth]{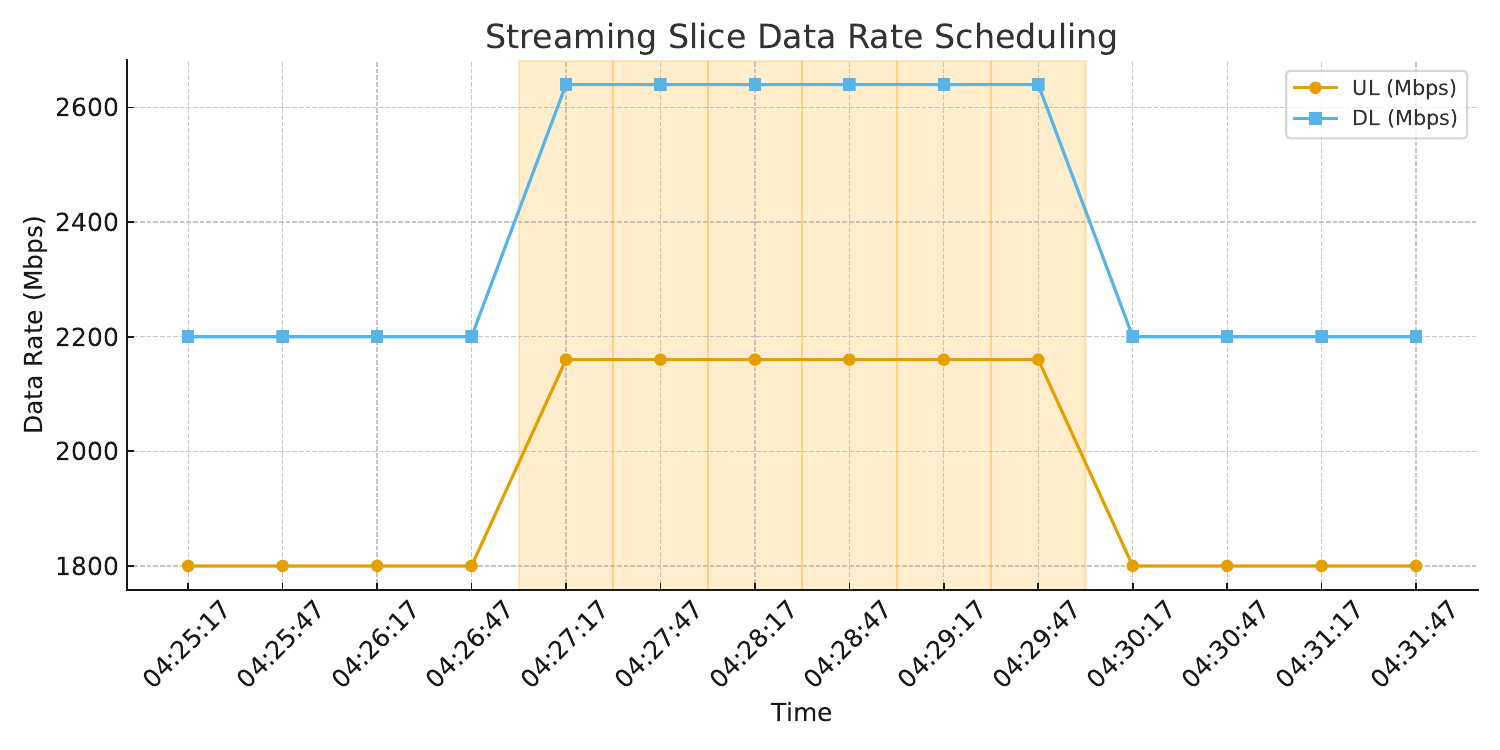}
	\caption{Policy change for scheduled data rate increase in streaming slice.}
	\label{fig:res2}
\end{figure}

\begin{tcolorbox}[colback=blue!20!white, colframe=blue, title=Use Case 2: Slice policy scheduling]
Increase the data rate for the 'streaming' slice by 20\% from 4:27 PM until 4:30 PM on weekdays.
\end{tcolorbox}
For the second intent, the network operator asked IntAgent to schedule an increase in the data rate for a specific slice between 4:27 PM and 4:30 PM on weekdays. This short and precise interval was deliberately chosen to highlight the system’s ability to enforce fine-grained, time-critical policies. Such timing reflects practical scenarios where sudden surges in demand occur, such as the release of popular streaming content or micro-peak events. The agent generated a plan as shown below; then, after checking the feasibility, it contacted the monitoring manager to schedule a policy change to increase the AMBR (Aggregate Maximum Bit Rate) for all 'streaming' slice sessions. As shown in \ref{fig:res2}, the agent successfully scheduled and applied the change as the user requested.

\begin{tcolorbox}[colback=orange!10!white, colframe=orange, title=\textbf{Use Case 2: Plan}]
1. List collections to locate slice info. \\
2. Search for streaming slice config. \\
3. Check feasibility of +20\% increase. \\
4. Compute new data rate. \\
5. Apply update with schedule (4:27–4:30, weekdays). \\
6. Verify new config. \\
7. Provide a final answer to the user.
\end{tcolorbox}

\subsection{IntAgent Limitations}
Despite the promising outcomes demonstrated by the proposed solution and its tested use cases, certain limitations exist.
Firstly, although we implemented critical thinking in the observations step, the LLM hallucinations were happening regularly, and this could be due to the size of the chosen LLM model.
Moreover, we observed that the agent's complex plan-(think-act-observe) cycle results in numerous additional steps even for basic intents like carrying out straightforward tasks.

\section{CONCLUSION AND FUTURE WORK}
In this paper, we introduced IntAgent, an LLM-based agent framework that processes network operator intents by integrating with the 3GPP NWDAF. The IntAgent system uses its custom-designed intent tools engine, which operates on real-time standardized analytics to convert operator intentions into specific operational commands. The system demonstrated its effectiveness through tested use cases, which showed how machine learning prediction and scheduled policy were applied successfully by giving only a high-level intent. Future research will focus on fine-tuning the LLM using domain-specific networking data, which should decrease hallucinations while enhancing its operational effectiveness using an improved system prompt. We also plan to conduct a study that compares the performance of various LLM models. Finally, our long-term vision is to expand IntAgent into a collaborative multi-agent system capable of autonomously managing complex, cross-domain network intents.

\label{sec:conc}


\bibliographystyle{IEEEtran}
\bibliography{biblio}
\end{document}